\documentclass[11pt]{raa}

\usepackage{graphicx,times}
\usepackage{natbib}
\usepackage{amssymb,amsmath}
\usepackage[pagebackref=true]{hyperref}
\bibpunct{(}{)}{;}{a}{}{,}
\usepackage{amsmath}	
\usepackage{tikz}
\usepackage{hyperref}

\begin{document}
\usetikzlibrary{shapes.geometric, arrows}

\def\Ntheory{N_{\rm th}}
\def\Nobs{N_{\rm obs}}
\def\probof#1{\mathrm{P}\left(#1\right)}
\def\Ptail{\probof{\Ntheory \geq \Nobs}}
\def\Ptheory{\probof{\Ntheory}}
\def\Pobs{\probof{\Nobs}}
\def\plotone#1{\includegraphics[width=0.8\textwidth]{#1}}


\title{Quantifying the tension between cosmological models and JWST red candidate massive galaxies}

\volnopage{ {\bf 20XX} Vol.\ {\bf X} No. {\bf XX}, 000--000}
   
\setcounter{page}{1}

\author{Jun-Chao Wang\inst{1}, Zhi-Qi Huang\inst{*1,2}, Lu Huang\inst{3}, Jianqi Liu\inst{1}}

\institute{
    School of Physics and Astronomy, Sun Yat-sen University, 2 Daxue Road, Tangjia, Zhuhai, 519082, China\\
    \and
    CSST Science Center for the Guangdong-Hong Kong-Macau Greater Bay Area, Zhuhai, 519082, China; {\it huangzhq25@mail.sysu.edu.cn}\\
    \and
    CAS Key Laboratory of Theoretical Physics, Institute of Theoretical Physics, Chinese Academy of Sciences, Beijing, 100190, China \\    
    \vs \no
    {\small Received 20XX Month Day; accepted 20XX Month Day}
}

\abstract{We develop a Python tool to estimate the tail distribution of the number of dark matter halos beyond a mass threshold and in a given volume in a light-cone. The code is based on the extended Press-Schechter model and is computationally efficient, typically taking a few seconds on a personal laptop for a given set of cosmological parameters. The high efficiency of the code allows a quick estimation of the tension between cosmological models and the red candidate massive galaxies released by the James Webb Space Telescope, as well as scanning the theory space with the Markov Chain Monte Carlo method. As an example application, we use the tool to study the cosmological implication of the candidate galaxies presented in \citet{labbe2023population}. The standard $\Lambda$ cold dark matter ($\Lambda$CDM) model is well consistent with the data if the star formation efficiency can reach $\sim 0.3$ at high redshift. For a low star formation efficiency $\epsilon \sim 0.1$, $\Lambda$CDM model is disfavored at $\sim 2\sigma$-$3\sigma$ confidence level.
}

\authorrunning{J.-C. Wang et al. }            
\titlerunning{Tension between cosmological models with JWST}

\maketitle


\section{Introduction} \label{sec:intro}

In the standard $\Lambda$ cold dark matter ($\Lambda$CDM) picture, baryon decouples from radiation after recombination and tracks the gravitational potential of dark matter. Before the formation of early galaxies, baryon and dark matter are well mixed. The fraction of baryonic mass in a massive dark matter halo therefore approximately equals to the cosmological mean $f_b \equiv \Omega_b/\Omega_m$, where $\Omega_b$ and $\Omega_m$ are the baryon and matter abundance parameters, respectively. If the fraction of stellar mass in baryonic matter, namely the star formation efficiency (SFE) $\epsilon$ is known, we can connect the stellar mass $M_\star$ to the halo mass $M_{\rm halo}$ via $M_\star  = \epsilon f_b M_{\rm halo}$. This provides a way to test cosmology by estimating stellar masses of massive galaxies in a given volume. In particular, at high redshift where massive halos are predicted to be very rare, detection of very massive galaxies may pose a stringent constraint on cosmology. The red candidate massive galaxies released by the James Webb Space Telescope (JWST), for instance, have inspired a debate whether a beyond-$\Lambda$CDM cosmology is on demand~\citep{labbe2023population, boylan-kolchin2023stress, lovell2023extreme, haslbauer2022has,  chen2023massive, ferrara2023on, qin2023implications, wang2023modeling}. A plenty of beyond-$\Lambda$CDM models including alternative dark energy~\citep{wang2023exploring, adil2023dark, menci2022high}, non-standard dark matter~\citep{bird2023high, lin2023implications, maio2023jwst, dayal2023warm}, cosmic string~\citep{jiao2023early}, primordial black holes~\citep{yuan2023rapidly, hutsi2023did, gouttenoire2023scrutinizing, guo2023footprints, huang2023supermassive}, and many others~\citep{melia2023cosmic, john2023cosmological, lovyagin2022cosmological, lei2023black} have been confronted with the JWST high-redshift candidate galaxies. Modification of primordial conditions may also lead to an overabundance of dark matter halos and hence more massive galaxies at high redshift~\citep{huang2019high, parashari2023primordial, padmanabhan2023alleviating, keller2023can, forconi2023do, su2023an, wang2023jwst, tkachev2023excess}.

The major obstacle to an accurate comparison between a theory and observed galaxies is the uncertainty in the star formation efficiency $\epsilon$ at high redshift. Galaxy formation models and low-redshift observation typically suggest $\epsilon \lesssim 0.2$~\citep{guo2020further, boylan-kolchin2023stress}, though $\epsilon$ at high redshift may be significant different~\citep{zhang2022massive, qin2023implications}. Because the UV radiation from star formation can ionize the neutral hydrogen at the reionization epoch, a larger star formation efficiency tends to accelerate the reionization process and finish it at an earlier time. Observations of the Lyman-$\alpha$ emitter luminosity function suggest that reionization is still on-going at $z\gtrsim 7$~\citep{wold2022large, goto2021silverrush}, consistent with the $\epsilon\lesssim 0.2$ picture~\citep{minoda2023reionization,mobina2023a}. It is however difficult to derive an upper-bound of $\epsilon$ solely from the reionization history, due to the degeneracy between $\epsilon$ and other astrophysical parameters.

The common practice is then to study the cosmological implication for an assumed star formation efficiency. Although most of the studies qualitatively agree, quantitative discrepancies still exist. For example, while \cite{boylan-kolchin2023stress} and \citet{chen2023massive} find that $\epsilon \sim 0.5$ suffices to explain the \cite{labbe2023population} data in the standard $\Lambda$CDM cosmology, \citet{menci2022high} that uses the same data claims that $\Lambda$CDM is ruled out at $2\sigma$ level even with $\epsilon = 1$. This is not only due to different choices of mass cut and volume cut, but also because it is hard to give an accurate description of the systematic errors of redshift and stellar mass obtained with spectral energy distribution (SED) fitting. By using seven individual redshift and stellar mass measurements with different SED-fitting codes/templates, \citet{labbe2023population} estimated extremes of stellar mass and redshift of each candidate galaxy. These extremes however are insufficient for quantitative analysis, which requires precise knowledge of the probability density functions (PDFs) of the systematic errors. 

The summary statistics of the statistical errors of stellar mass are often represented as $M^{+\Delta M_{\rm sup}}_{-\Delta M_{\rm inf}}$, where $M$ is the median mass. The upper and lower errors ($\Delta M_{\rm sup}$ and $\Delta M_{\rm inf}$) are defined with $16$th and $84$th percentiles. The errors in stellar mass, and similarly in redshift if only photometry information is available, are typically asymmetric and difficult to model analytically. Accurate observation-theory comparison usually requires an end-to-end comparison between high-resolution simulations and observations. On the other hand, while the summary statistics (and the extremes from different SED-fitting methods, if available) contain limited information, they are important science products of many galaxy surveys. The aim of the present work is to develop a tool that performs a preliminary scan of the theory space based on these simple products, before planning subsequent spectroscopic observation and running costly simulations. 

This paper is organized as follows. Section~\ref{sec:model} describes the physical model of the abundance of high-redshift massive galaxies and the workflow of observation-theory comparison. Section~\ref{sec:application} uses the tool to analyze the cosmological implication of the candidate galaxies in \citet{labbe2023population}. Section~\ref{sec:conc} concludes.

\section{Algorithm of the Tool } \label{sec:model}

\tikzstyle{standardbox} = [rectangle, rounded corners, minimum width=2.6cm, minimum height=1cm,text centered, draw=black, fill=orange!30]
\tikzstyle{arrow} = [thick,->,>=stealth]

\begin{figure*}
\centering
\begin{tikzpicture}[node distance=1.5cm]
\node(sfe)[standardbox]{SFE};
\node(galmcut)[standardbox, below of=sfe]{galaxy $M_{\rm  \star, cut}$};
\node(cosmology)[standardbox, below of=galmcut]{cosmology};
\node(fb)[standardbox, right of=galmcut, xshift=1.8cm]{$f_b=\Omega_b/\Omega_m$};
\node(volcut)[standardbox, below of=cosmology]{volume cut};
\node(data)[standardbox, below of=volcut]{observed galaxies};
\node(statmodel)[standardbox, below of=data]{model of error PDFs};
\node(halomcut)[standardbox, right of=sfe, xshift=5.2cm]{halo $M_{\rm cut}$};
\node(PNtheory)[standardbox, right of=cosmology, xshift=8.4cm]{$\Ptheory$};
\node(PNobs)[standardbox, right of=data, xshift=8.4cm]{$\Pobs$};
\node(Ptail)[standardbox, right of=volcut, xshift=11.6cm]{$\Ptail$};
\draw[arrow](sfe)--(halomcut);
\draw[arrow](galmcut)--(halomcut);
\draw[arrow](cosmology)--(fb);
\draw[arrow](fb)--(halomcut);
\draw[arrow](halomcut)-|(PNtheory);
\draw[arrow](cosmology)--(PNtheory);
\draw[arrow](volcut)--(PNtheory);
\draw[arrow](volcut)--(PNobs);
\draw[arrow](data)--(PNobs);
\draw[arrow](statmodel)-|(PNobs);
\draw[arrow](PNtheory)-|(Ptail);
\draw[arrow](PNobs)-|(Ptail);
\end{tikzpicture}
\caption{Workflow of observation-theory comparison \label{fig:workflow}}    
\end{figure*}
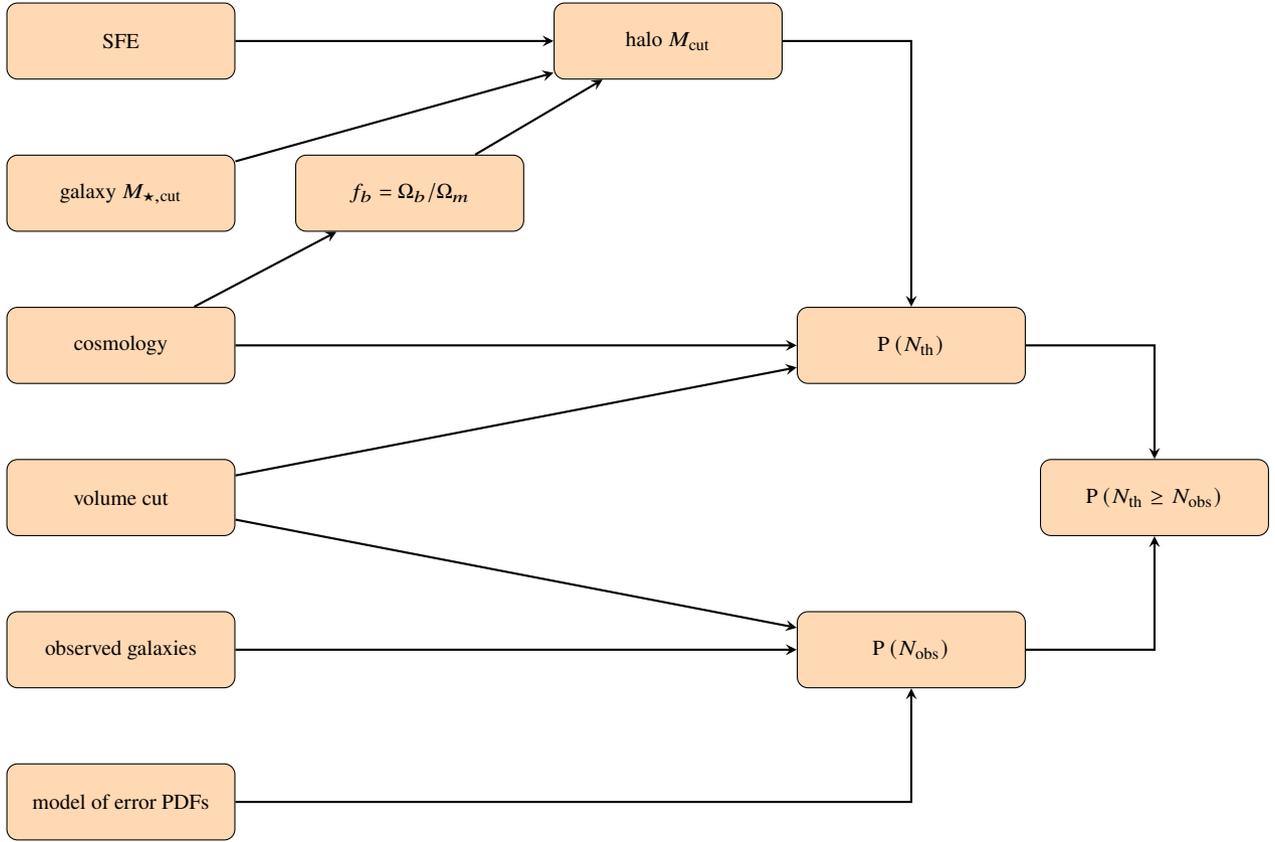

Figure~\ref{fig:workflow} summarizes our algorithm to do an observation-theory comparison. We consider galaxies beyond a stellar-mass threshold ($M_\star > M_{\rm \star, cut}$) and in a fixed comoving volume defined by the sky coverage $f_{\rm sky}$ and redshift range ($z_{\min} < z < z_{\max}$). Because both the stellar mass and redshift from SED-fitting have significant uncertainties, whether an observed galaxy is above the stellar-mass threshold and in the volume is probabilistic.  The total observed galaxies above the stellar-mass threshold and in the volume, denoted as $\Nobs$, is therefore a random variable. Due to the cosmic variance and the Poisson shot noise, the theoretical prediction of the total number of galaxies above the stellar-mass threshold and in the volume, denoted as $\Ntheory$, is also probabilistic.

A model is ruled out when and only when $\Ntheory < \Nobs$. The tension between a cosmological model and the observed galaxies is therefore quantified by the probability $\probof{\Ntheory < \Nobs}$, or more conveniently, the smallness of $\Ptail = 1- \probof{\Ntheory < \Nobs}$. For instance, $\Ptail =0.01$ indicates that the model under consideration is ruled at $99\%$ confidence level. Here and in what follows, $\probof{\cdot}$ stands for probability (for discrete variable or an event) or probability density function (for continuous variable).

For a given cosmology, $\Ptail$ is evaluated in three steps.
\begin{itemize}
 \item[1.]{For $\Nobs=0, 1, 2,\ldots$ compute the distribution function $\probof{\Nobs}$ from summary statistics of observed galaxies and user-specified assumptions about the functional form of probability density functions of stellar mass and redshift.}
 \item[2.]{For $\Ntheory = 0,1, 2,\ldots$ compute the distribution function $\probof{\Ntheory}$ from the cosmological model and a user-specified star formation efficiency.}
 \item[3.]{Compute $$\Ptail  = \sum_{\Nobs=0}^\infty 
 \probof{\Nobs} \sum_{\Ntheory=\Nobs}^\infty \probof{\Ntheory}.$$}
\end{itemize}

Explicit evaluation of $\Ptail$ requires perfect knowledge about the PDFs of statistical/systematic errors of the stellar masses and redshifts of the candidate galaxies. While the PDFs of statistical errors can be recovered by going back to the SED-fitting process, it is not easy to give an accurate description of the PDFs of systematic errors, which often dominate the error budget. Without further knowledge about the systematic errors, we simply model the PDFs with a few generic distribution functions: skew normal, skew triangular, and skew rectangular, whose definitions are given in the Appendix. The code is organized in such a way that the users can easily add their own distribution functions. We also allow the users to turn off systematic errors by using Dirac delta distribution. In the case when systematic errors are given in the form of extremes from different SED-fitting methods, we match the extremes with exact upper/lower bounds of skew triangular/rectangular distributions or $2\sigma$ bounds of skew normal distribution. Finally, the stellar mass $M_\star$ and redshift $z$ from different SED-fitting methods are typically positively correlated. We test the impact of $M_\star$-$z$ correlation by forcing $(M_\star - M_{\rm \star, median})(z-z_{\rm median})\geq 0$ in the joint distribution of $M_\star$ and $z$, and find that the correlation does not make a qualitative difference.

The evaluation of $\Ptheory$ involves the extended Press-Schechter ellipsoidal collapse model~\citep{press1974formation,sheth2001ellipsoidal,sheth2002excursion}, where the comoving halo number density per mass interval, namely the halo mass function (at a fixed redshift $z$) is given by
\begin{equation} \label{equ:hmf}
  \frac{\mathrm{d}n}{\mathrm{d}M} = -f(\nu) \frac{\rho_m}{M}\frac{\mathrm{d}\ln \sigma}{\mathrm{d}M}.
\end{equation}
The $\rho_{m}$ on the right-hand side is the comoving average background matter density and $\sigma$ is the mass fluctuation at scale $R=\left(\frac{3M}{4\pi \rho_m}\right)^{1/3}$, given by
\begin{equation}
    \sigma^2=\frac{1}{2\pi ^2}\int_0^{\infty}k^2 \mathrm{P}_m(k,z)\mathrm{W}^2(kR)\,\mathrm{d}k,
\end{equation}
where $k$ is the comoving wavenumber, and $\mathrm{P}_m(k, z)$ is the linear matter power spectrum at redshift $z$. For the filter function $\mathrm{W}(kR)$, we take the form of top-hat
\begin{equation}
  \mathrm{W}(x)=\frac{3(\sin x-x\cos x)}{x^3}.
\end{equation}
The simulation calibrated factor $f(\nu)$ is given by
\begin{equation}
  f(\nu)=2A\left[1+\frac{1}{(a\nu^2)^q}\right]\left(\frac{a\nu^2}{2\pi}\right)^{1/2}\exp (-\frac{a\nu^2}{2}),
\end{equation}
with $A=0.322$, $a=0.707$, $q=0.3$ and $\nu=\frac{\delta_c}{\sigma(R,z)}$, where parameter $\delta_c=1.686$ corresponds to the critical linear overdensity. We ignore the tiny difference in $\delta_c$ for different cosmological models, as we are only interested in the models that are close to the concordance $\Lambda$CDM model. 

For a massive galaxy with stellar mass $M_\star > M_{\rm \star,cut}$, the mass of its host halo must exceed $M_{\rm cut} = M_{\rm \star,cut}/(\epsilon f_b)$. The halo mass function Eq.~\eqref{equ:hmf} allows to compute the ensemble average of the mean number of such halos
\begin{equation}
  \langle\Ntheory\rangle=4\pi f_{\rm sky}\int_{M_{\rm cut}}^{\infty}\mathrm{d}M \int_{z_{\min}}^{z_{\max}} \frac{\mathrm{d}n}{\mathrm{d}M} \frac{\mathrm{d}V}{\mathrm{d}z\mathrm{d}\Omega} \,\mathrm{d}z, \label{eq:Nbar}
\end{equation}
where $\frac{\mathrm{d}V}{\mathrm{d}z\mathrm{d}\Omega}$ is comoving volume per redshift interval per solid angle. Here we have equalized the number of massive galaxies and the number of their host halos, assuming each massive halo has a central galaxy.

For the purpose of using rare objects to study cosmological tensions, we are only interested in the $\langle\Ntheory\rangle\lesssim O(1)$ case, where $\Ntheory$ approximately follows a Poisson distribution
\begin{equation}
\Ptheory \sim e^{-\lambda}\frac{\lambda^{\Ntheory}}{\Ntheory!}, \label{eq:Ndis_approx}
\end{equation}
where $\lambda \approx \langle \Ntheory \rangle$ is given in Eq.~\eqref{eq:Nbar}. 

To obtain a more accurate result, we integrate $\lambda$ on the right-hand side of Eq.~\eqref{eq:Ndis_approx} over a Gaussian distribution, giving
\begin{equation}
\Ptheory = \int_{-\infty}^\infty \frac{1}{\sqrt{2\pi}\sigma_\lambda}e^{-\frac{(\lambda-\langle \Ntheory\rangle)^2}{2\sigma_\lambda^2}}  e^{-\lambda}\frac{\lambda^{\Ntheory}}{\Ntheory!}\,\mathrm{d}\lambda. \label{eq:Ndis_exact}
\end{equation}
The beyond-Poisson cosmic variance from the large scale structure, $\sigma_\lambda$, can be read out from the online emulator at \url{https://www.ph.unimelb.edu.au/~mtrenti/cvc/CosmicVariance.html}~\citep{trenti2008cosmic}. For a pencil-like survey volume, we find the cosmic variance approximately equal to the product of the linear result, which equals to the product of linear halo bias and matter cosmic variance, and a fudge factor $\approx 1.4$ that accounts for the nonlinear correction. Since the cosmic-variance correction is a subdominant effect\footnote{Our result does not contradict that of \citet{chen2023massive}, which claims that cosmic variance taken from their simulations is a dominant factor. Our ``cosmic variance'' here, as well as that defined in \citet{trenti2008cosmic}, refers to the variance due to power spectrum uncertainties caused by finite sampling and does not include Poisson shot noise, while ``cosmic variance'' taken from simulations include both effects.  }, this approximation suffices and allows us to make our tool self-contained.

The algorithm is realized with a Python tool publicly available at \url{http://zhiqihuang.top/codes/massivehalo.htm}.

\section{Applying to JWST data} \label{sec:application}

We now apply the tool to the 13 candidate galaxies in \citet{labbe2023population}. The summary statistics of stellar mass and redshift are given in the Extended Data Table 2 of \cite{labbe2023population}, as well as in the online repository given in Sec.~\ref{sec:data}. Stellar mass and redshift values include two uncertainties: $\pm (\text{ran}) \pm (\text{sys})$, where the random uncertainty ($\text{ran}$) corresponds to the $16$th and $84$th percentiles of the combined posterior distribution, and the systematic uncertainty ($\text{sys}$) corresponds to the extreme values from seven different methods (EAZY, Prospector, and Bagpipes with 5 variations, including dust, SFH, age prior, and SNR limit). In this study, we update the data of three candidate galaxies (L23-13050, L23-35300 and L23-39575) with the information from spectroscopic follow-up observations~\citep{kocevski2023hidden,perez2023ceers, fujimoto2023ceers}. The updated data is also available in the online repositary given in Sec.~\ref{sec:data}.

\begin{figure}
  \centering
\plotone{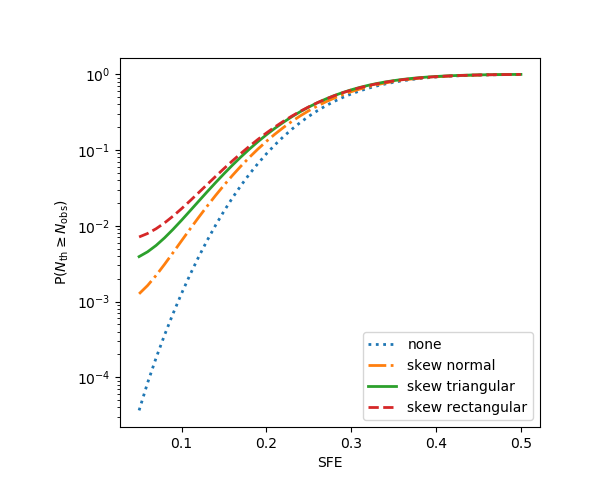} 
\caption{Dependence of $\Ptail$ on the star formation efficiency and the model of the PDF of systematic errors.\label{fig:eps}}
\end{figure}

Unless otherwise specified, we choose the stellar-mass threshold to be $M_{\rm \star,cut} = 10^{10}M_\odot$ and the redshift range to be $z_{\min}=7$ and $z_{\rm max}=10$, and we assume the statistical errors follow a skew normal distribution. The data covers a sky area $38\,\mathrm{arcmin}^2$, which translates to $f_{\rm sky}=2.56\times 10^{-7}$. We take the cosmology to be the Planck best-fit $\Lambda$CDM model~\citep{aghanim2020planck} with $\Omega_m=0.3158$, $\Omega_b=0.049389$, primordial spectrum index $n_s=0.96605$, matter fluctuation parameter $\sigma_8=0.812$, and Hubble constant $H_0=67.32\,\mathrm{km/s/Mpc}$. Figure~\ref{fig:eps} shows the dependence of $\Ptail$ on the star formation efficiency and the model of the PDFs of systematic errors. For $\epsilon\gtrsim 0.3$ that is fairly possible at the early epoch of galaxy formation~\citep{zhang2022massive, qin2023implications}, the data is well consistent with $\Lambda$CDM model even when the systematic errors are ignored (dotted blue line in Figure~\ref{fig:eps}). For the most conservative $\epsilon \sim 0.1$ case, $\Lambda$CDM model is slightly disfavored by the data. For different forms of the PDFs of the systematic errors, the statistical significance of the tension varies between $\sim 2\sigma$ and $\sim 3\sigma$.

\begin{figure}
  \centering
\plotone{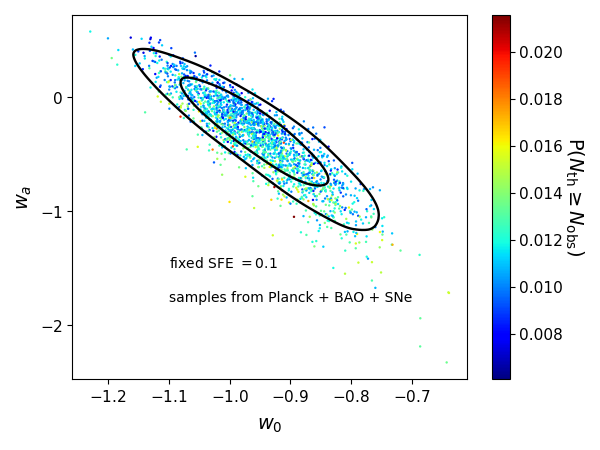}
\caption{$\Ptail$ for $w_0$-$w_a$ cosmology. The star formation efficiency is fixed to be $0.1$. Systematic errors are assumed to follow skew triangular distribution. For each point, the cosmological parameters are drawn from thinned Planck+BAO+SNe Markov chains. The solid black lines are the marginalized $68.3\%$ and $95.4\%$ confidence contours for the Planck+BAO+SNe likelihood.\label{fig:w0wa}}
\end{figure}

It has been shown that dynamic dark energy model with equation of state $w(a) = w_0 + w_a(1-a)$~\citep{Chevallier01,Linder03}, where $a$ is the scale factor, may alleviate the tension between cosmology and the \citet{labbe2023population} data. Since the tension is only significant when the star formation efficiency $\epsilon$ is $\lesssim 0.1$, we hereafter fix $\epsilon=0.1$ and investigate whether the dynamic dark energy model is well consistent with the data. For simplicity, we also assume that the systematic errors follow the skew triangular distribution. We take cosmological parameters from the Markov chains that are sampled with Planck+BAO+SNe likelihood. Here Planck refers to the TTTEEE+lowl+LowE likelihood of the cosmic microwave background maps measured by the Planck satellite~\citep{aghanim2020planck}; BAO refers to the baryon acoustic oscillations data that are summarized in ~\citet{Alam2017clustering}; SNe stands for the Pantheon supernova catalog~\citep{Scolnic2018complete}. We download the chains from Planck Legacy Archive (\url{https://pla.esac.esa.int/pla/}) and, for better visual effect, thin the chains by a factor of 5. For each set of cosmological parameters of the thinned chains, we compute $\Ptail$ value and show it in Figure~\ref{fig:w0wa} as a dot, whose color represents the value of $\Ptail$ and coordinates indicate the dark energy parameters $(w_0, w_a)$. The variation of the dynamic dark energy equation of state in the range that is allowed by Planck+BAO+SNe does lead to a variation of $\Ptail$, however, not at a very significant level that would make the result qualitatively different.

\section{Discussion and Conclusions } \label{sec:conc}

In this work, we present and make publicly available an efficient numeric tool that can quickly estimate the tension between cosmological models and observed high-redshift massive galaxies. Our method is based on the statistics of the total numbers of massive halos, whereas the previous works are all based on the statistics of the cumulative stellar mass $\rho_\star(>M_\star)$~\citep{labbe2023population,menci2022high,boylan-kolchin2023stress,wang2023exploring,chen2023massive}. The advantage of using the number of halos as a fundamental variable is that the dominating Poisson shot noise contribution can be explicitly formulated.

We update the parameters of some of the candidate galaxies in \citet{labbe2023population} with recent spectroscopic follow-up observations. For a low star formation efficiency $\epsilon \sim 0.1$, the data remains in $\gtrsim 2\sigma$ tension with the concordance $\Lambda$CDM cosmology, and the tension is insensitive to the dark energy equation of state and other cosmological parameters, provided that the cosmological parameters are confined by the Planck+BAO+SNe likelihood. We also verify that the updated spectroscopic redshift information slightly reduces the tension between $\Lambda$CDM model and the JWST candidate galaxies. For instance, assuming a skew normal distribution of both systematic and statistical errors and star formation efficiency $\epsilon=0.1$, we find Planck best-fit $\Lambda$CDM model is ruled out at $2.59\sigma$ level with the original data from \citet{labbe2023population} and $2.49\sigma$ level with the updated data.

Although the algorithm in Figure~\ref{fig:workflow} captures the main physics, there are subtleties beyond the scope of the present work. First, it is known that the extended Press-Schechter halo mass functions at high redshift are 20\%-50\% higher than those obtained from N-body simulations~\citep{reed2003evolution,despali2016universality,shirasaki2021virial,wang2022ultramarine}. Secondly, backsplash halos, whose dark matter has left the host halo while its gas may remain in the host halo~\citep{balogh2000origin,wang2009distribution,wetzel2014galaxy,diemer2021flybys,wang2023dissect}, may make the interpretation of star formation efficiency ambiguous in the real world and may alter the statistical significance of the cosmological tension~\citep{chen2023massive}. Finally, we have ignored catastrophic redshift errors which, despite being rare, might have a significant impact on the counting of rare objects. We leave further studies on these nontrivial subjects as our future work.

\normalem
\begin{acknowledgements}

This work is supported by National key R\&D  Program of China (Grant No. 2020YFC2201600), the National Natural Science Foundation of China (NSFC) under Grant No. 12073088, and National SKA Program of China No. 2020SKA0110402.

\end{acknowledgements}



\appendix

\section{Modelling the Distribution Functions of Stellar Mass and Redshift}

Here we introduce the distribution functions that are built in the tool. 

\subsection{skew normal distribution}

The standardized skew normal distribution with parameter $a$ and of a random variable $x$ is given by the PDF
\begin{equation}
\mathcal{N}(x; a) = \frac{1}{\sqrt{2\pi}}e^{-x^2/2}\left[1+\mathrm{erf}\,\left(ax/\sqrt{2}\right)\right].\label{eq:ssknew}
\end{equation}
In the general case, the skew normal distribution is obtained by rescaling and translating $x$,
\begin{equation}
\mathrm{N}(x; a, \zeta, \omega) = \mathcal{N}\left(\frac{x-\zeta}{\omega}; a\right). \label{eq:skewnorm}
\end{equation}
For a given summary statistics, the parameters $a, \zeta, \omega$ can be fixed by matching the $16$th and $84$th percentiles and the median. The skew normal distribution \eqref{eq:skewnorm} cannot describe a very skewed distribution with the ratio between two errors exceeding $1.55$. In this extreme case, we replace the linear transformation on $x$ in \eqref{eq:ssknew} with a nonlinear one. Since this procedure is quite complicated and does not have much to do with the scientific result in the present work, we refer interested readers to the publicly available source code \url{http://zhiqihuang.top/codes/fitskew.py}.

\subsection{skew triangular and skew rectangular}

For parameters $a>0$, $b>0$ and $\mu$, the skew triangular distribution of a random variable $x$ is defined as
\begin{equation}
\mathrm{T}(x; \mu, a, b) = \left\{
\begin{array}{ll}
\frac{x-\mu+a}{a^2}, &  \text{ if } \mu-a< x \le \mu ; \\
\frac{\mu+b-x}{b^2}, &  \text{ if } \mu < x < \mu + b ; \\
0, & \text{ else.}
\end{array}
\right.
\end{equation}
The skew rectangular distribution is defined as
\begin{equation}
\mathrm{R}(x; \mu, a, b) = \left\{
\begin{array}{ll}
\frac{1}{2a}, &  \text{ if } \mu-a< x \le \mu ; \\
\frac{1}{2b}, &  \text{ if } \mu < x < \mu + b ; \\
0, & \text{ else.}
\end{array}
\right.
\end{equation}


\begin{thebibliography}{60}
\providecommand\natexlab[1]{#1}
\providecommand\JournalTitle[1]{#1}

\bibitem[{Adil} {et~al.}(2023)]{adil2023dark}
{Adil}, S.~A., {Mukhopadhyay}, U., {Sen}, A.~A., \& {Vagnozzi}, S. 2023, arXiv
  e-prints, arXiv:2307.12763

\bibitem[Aghanim {et~al.}(2020)]{aghanim2020planck}
Aghanim, N., Akrami, Y., Ashdown, M., {et~al.} 2020, Astronomy \& Astrophysics,
  641, A6

\bibitem[{Alam} {et~al.}(2017)]{Alam2017clustering}
{Alam}, S., {Ata}, M., {Bailey}, S., {et~al.} 2017, \mnras, 470, 2617

\bibitem[{Balogh} {et~al.}(2000)]{balogh2000origin}
{Balogh}, M.~L., {Navarro}, J.~F., \& {Morris}, S.~L. 2000, \apj, 540, 113

\bibitem[{Bird} {et~al.}(2023)]{bird2023high}
{Bird}, S., {Chang}, C.-F., {Cui}, Y., \& {Yang}, D. 2023, arXiv e-prints,
  arXiv:2307.10302

\bibitem[{Boylan-Kolchin}(2023)]{boylan-kolchin2023stress}
{Boylan-Kolchin}, M. 2023, Nature Astronomy, 7, 731

\bibitem[{Chen} {et~al.}(2023)]{chen2023massive}
{Chen}, Y., {Mo}, H.~J., \& {Wang}, K. 2023, arXiv e-prints, arXiv:2304.13890

\bibitem[{Chevallier} \& {Polarski}(2001)]{Chevallier01}
{Chevallier}, M., \& {Polarski}, D. 2001, International Journal of Modern
  Physics D, 10, 213

\bibitem[{Dayal} \& {Giri}(2023)]{dayal2023warm}
{Dayal}, P., \& {Giri}, S.~K. 2023, arXiv e-prints, arXiv:2303.14239

\bibitem[{Despali} {et~al.}(2016)]{despali2016universality}
{Despali}, G., {Giocoli}, C., {Angulo}, R.~E., {et~al.} 2016, \mnras, 456, 2486

\bibitem[{Diemer}(2021)]{diemer2021flybys}
{Diemer}, B. 2021, \apj, 909, 112

\bibitem[{Ferrara} {et~al.}(2023)]{ferrara2023on}
{Ferrara}, A., {Pallottini}, A., \& {Dayal}, P. 2023, \mnras, 522, 3986

\bibitem[{Forconi} {et~al.}(2023)]{forconi2023do}
{Forconi}, M., {Ruchika}, {Melchiorri}, A., {Mena}, O., \& {Menci}, N. 2023,
  arXiv e-prints, arXiv:2306.07781

\bibitem[Fujimoto {et~al.}(2023)]{fujimoto2023ceers}
Fujimoto, S., Haro, P.~A., Dickinson, M., {et~al.} 2023, The Astrophysical
  Journal Letters, 949, L25

\bibitem[{Goto} {et~al.}(2021)]{goto2021silverrush}
{Goto}, H., {Shimasaku}, K., {Yamanaka}, S., {et~al.} 2021, \apj, 923, 229

\bibitem[{Gouttenoire} {et~al.}(2023)]{gouttenoire2023scrutinizing}
{Gouttenoire}, Y., {Trifinopoulos}, S., {Valogiannis}, G., \& {Vanvlasselaer},
  M. 2023, arXiv e-prints, arXiv:2307.01457

\bibitem[{Guo} {et~al.}(2020)]{guo2020further}
{Guo}, Q., {Hu}, H., {Zheng}, Z., {et~al.} 2020, Nature Astronomy, 4, 246

\bibitem[{Guo} {et~al.}(2023)]{guo2023footprints}
{Guo}, S.-Y., {Khlopov}, M., {Liu}, X., {et~al.} 2023, arXiv e-prints,
  arXiv:2306.17022

\bibitem[{Haslbauer} {et~al.}(2022)]{haslbauer2022has}
{Haslbauer}, M., {Kroupa}, P., {Zonoozi}, A.~H., \& {Haghi}, H. 2022, \apjl,
  939, L31

\bibitem[{Huang} {et~al.}(2023)]{huang2023supermassive}
{Huang}, H.-L., {Cai}, Y., {Jiang}, J.-Q., {Zhang}, J., \& {Piao}, Y.-S. 2023,
  arXiv e-prints, arXiv:2306.17577

\bibitem[{Huang}(2019)]{huang2019high}
{Huang}, Z. 2019, \prd, 99, 103537

\bibitem[{H{\"u}tsi} {et~al.}(2023)]{hutsi2023did}
{H{\"u}tsi}, G., {Raidal}, M., {Urrutia}, J., {Vaskonen}, V., \& {Veerm{\"a}e},
  H. 2023, \prd, 107, 043502

\bibitem[{Jiao} {et~al.}(2023)]{jiao2023early}
{Jiao}, H., {Brandenberger}, R., \& {Refregier}, A. 2023, arXiv e-prints,
  arXiv:2304.06429

\bibitem[{John} \& {Babu Joseph}(2023)]{john2023cosmological}
{John}, M.~V., \& {Babu Joseph}, K. 2023, arXiv e-prints, arXiv:2306.04577

\bibitem[{Keller} {et~al.}(2023)]{keller2023can}
{Keller}, B.~W., {Munshi}, F., {Trebitsch}, M., \& {Tremmel}, M. 2023, \apjl,
  943, L28

\bibitem[Kocevski {et~al.}(2023)]{kocevski2023hidden}
Kocevski, D.~D., Onoue, M., Inayoshi, K., {et~al.} 2023, arXiv preprint
  arXiv:2302.00012

\bibitem[Labb{\'e} {et~al.}(2023)]{labbe2023population}
Labb{\'e}, I., van Dokkum, P., Nelson, E., {et~al.} 2023, Nature, 616, 266

\bibitem[{Lei} {et~al.}(2023)]{lei2023black}
{Lei}, L., {Zu}, L., {Yuan}, G.-W., {et~al.} 2023, arXiv e-prints,
  arXiv:2305.03408

\bibitem[{Lin} {et~al.}(2023)]{lin2023implications}
{Lin}, H., {Gong}, Y., {Yue}, B., \& {Chen}, X. 2023, arXiv e-prints,
  arXiv:2306.05648

\bibitem[{Linder}(2003)]{Linder03}
{Linder}, E.~V. 2003, Physical Review Letters, 90, 091301

\bibitem[{Lovell} {et~al.}(2023)]{lovell2023extreme}
{Lovell}, C.~C., {Harrison}, I., {Harikane}, Y., {Tacchella}, S., \& {Wilkins},
  S.~M. 2023, \mnras, 518, 2511

\bibitem[{Lovyagin} {et~al.}(2022)]{lovyagin2022cosmological}
{Lovyagin}, N., {Raikov}, A., {Yershov}, V., \& {Lovyagin}, Y. 2022, Galaxies,
  10, 108

\bibitem[{Maio} \& {Viel}(2023)]{maio2023jwst}
{Maio}, U., \& {Viel}, M. 2023, \aap, 672, A71

\bibitem[{Melia}(2023)]{melia2023cosmic}
{Melia}, F. 2023, \mnras, 521, L85

\bibitem[{Menci} {et~al.}(2022)]{menci2022high}
{Menci}, N., {Castellano}, M., {Santini}, P., {et~al.} 2022, \apjl, 938, L5

\bibitem[{Minoda} {et~al.}(2023)]{minoda2023reionization}
{Minoda}, T., {Yoshiura}, S., \& {Takahashi}, T. 2023, arXiv e-prints,
  arXiv:2304.09474

\bibitem[{Mobina Hosseini} {et~al.}(2023)]{mobina2023a}
{Mobina Hosseini}, S., {Soleimanpour Salmasi}, B., {Sajad Tabasi}, S., \&
  {Firouzjaee}, J.~T. 2023, arXiv e-prints, arXiv:2306.12954

\bibitem[{Padmanabhan} \& {Loeb}(2023)]{padmanabhan2023alleviating}
{Padmanabhan}, H., \& {Loeb}, A. 2023, arXiv e-prints, arXiv:2306.04684

\bibitem[{Parashari} \& {Laha}(2023)]{parashari2023primordial}
{Parashari}, P., \& {Laha}, R. 2023, arXiv e-prints, arXiv:2305.00999

\bibitem[P{\'e}rez-Gonz{\'a}lez {et~al.}(2023)]{perez2023ceers}
P{\'e}rez-Gonz{\'a}lez, P.~G., Barro, G., Annunziatella, M., {et~al.} 2023, The
  Astrophysical Journal Letters, 946, L16

\bibitem[Press \& Schechter(1974)]{press1974formation}
Press, W.~H., \& Schechter, P. 1974, The Astrophysical Journal, 187, 425

\bibitem[{Qin} {et~al.}(2023)]{qin2023implications}
{Qin}, Y., {Balu}, S., \& {Wyithe}, J. S.~B. 2023, arXiv e-prints,
  arXiv:2305.17959

\bibitem[{Reed} {et~al.}(2003)]{reed2003evolution}
{Reed}, D., {Gardner}, J., {Quinn}, T., {et~al.} 2003, \mnras, 346, 565

\bibitem[{Scolnic} {et~al.}(2018)]{Scolnic2018complete}
{Scolnic}, D.~M., {Jones}, D.~O., {Rest}, A., {et~al.} 2018, \apj, 859, 101

\bibitem[Sheth {et~al.}(2001)]{sheth2001ellipsoidal}
Sheth, R.~K., Mo, H., \& Tormen, G. 2001, Monthly Notices of the Royal
  Astronomical Society, 323, 1

\bibitem[Sheth \& Tormen(2002)]{sheth2002excursion}
Sheth, R.~K., \& Tormen, G. 2002, Monthly Notices of the Royal Astronomical
  Society, 329, 61

\bibitem[{Shirasaki} {et~al.}(2021)]{shirasaki2021virial}
{Shirasaki}, M., {Ishiyama}, T., \& {Ando}, S. 2021, \apj, 922, 89

\bibitem[{Su} {et~al.}(2023)]{su2023an}
{Su}, B.-Y., {Li}, N., \& {Feng}, L. 2023, arXiv e-prints, arXiv:2306.05364

\bibitem[Tkachev {et~al.}(2023)]{tkachev2023excess}
Tkachev, M.~V., Pilipenko, S.~V., Mikheeva, E.~V., \& Lukash, V.~N. 2023,
  arXiv:2307.13774

\bibitem[{Trenti} \& {Stiavelli}(2008)]{trenti2008cosmic}
{Trenti}, M., \& {Stiavelli}, M. 2008, \apj, 676, 767

\bibitem[{Wang} \& {Liu}(2022)]{wang2023jwst}
{Wang}, D., \& {Liu}, Y. 2022, arXiv e-prints, arXiv:2301.00347

\bibitem[{Wang} {et~al.}(2009)]{wang2009distribution}
{Wang}, H., {Mo}, H.~J., \& {Jing}, Y.~P. 2009, \mnras, 396, 2249

\bibitem[{Wang} {et~al.}(2023{\natexlab{a}})]{wang2023dissect}
{Wang}, K., {Peng}, Y., \& {Chen}, Y. 2023{\natexlab{a}}, \mnras, 523, 1268

\bibitem[{Wang} {et~al.}(2023{\natexlab{b}})]{wang2023exploring}
{Wang}, P., {Su}, B.-Y., {Zu}, L., {Yang}, Y., \& {Feng}, L.
  2023{\natexlab{b}}, arXiv e-prints, arXiv:2307.11374

\bibitem[{Wang} {et~al.}(2022)]{wang2022ultramarine}
{Wang}, Q., {Gao}, L., \& {Meng}, C. 2022, \mnras, 517, 6004

\bibitem[{Wang} {et~al.}(2023{\natexlab{c}})]{wang2023modeling}
{Wang}, Y.-Y., {Lei}, L., {Yuan}, G.-W., \& {Fan}, Y.-Z. 2023{\natexlab{c}},
  arXiv e-prints, arXiv:2307.12487

\bibitem[{Wetzel} {et~al.}(2014)]{wetzel2014galaxy}
{Wetzel}, A.~R., {Tinker}, J.~L., {Conroy}, C., \& {van den Bosch}, F.~C. 2014,
  \mnras, 439, 2687

\bibitem[{Wold} {et~al.}(2022)]{wold2022large}
{Wold}, I. G.~B., {Malhotra}, S., {Rhoads}, J., {et~al.} 2022, \apj, 927, 36

\bibitem[{Yuan} {et~al.}(2023)]{yuan2023rapidly}
{Yuan}, G.-W., {Lei}, L., {Wang}, Y.-Z., {et~al.} 2023, arXiv e-prints,
  arXiv:2303.09391

\bibitem[{Zhang} {et~al.}(2022)]{zhang2022massive}
{Zhang}, Z., {Wang}, H., {Luo}, W., {et~al.} 2022, \aap, 663, A85

\end{thebibliography}

\end{document}